\documentclass[12pt]{article}

\usepackage{amsfonts, amsmath}

\newcommand{\be}{\begin{equation}} \newcommand{\ee}{\end{equation}}

\begin{document}
\title{The Universe as a Nonuniform Lattice in Finite-Volume  Hypercube.
          \\ I.Fundamental Definitions and Particular Features} \thispagestyle{empty}

\author{A.E.Shalyt-Margolin\hspace{1.5mm}\thanks
{Fax: (+375) 172 326075; e-mail: a.shalyt@mail.ru;alexm@hep.by}}
\date{}
\maketitle
 \vspace{-25pt}
{\footnotesize\noindent  National Center of Particles and High
Energy Physics, Bogdanovich Str. 153, Minsk 220040, Belarus\\
{\ttfamily{\footnotesize
}}

\rm\normalsize \vspace{0.5cm}
\begin{abstract}
In this paper a new small parameter associated with the density
matrix deformation (density pro-matrix)studied in previous works
of the author is introduced into the Generalized Quantum Mechanics
(GQM), i.e. quantum mechanics involving description of the Early
Universe. It is noted that this parameter has its counterpart in
the generalized statistical mechanics. Both parameters offer a
number of merits: they are dimensionless, varying over the
interval from 0 to 1/4 and assuming in this interval a discreet
series of values. Besides, their definitions contain all the
fundamental constants. These parameters are very small for the
conventional scales and temperatures, e.g. the value of the first
parameter is on the order of $\approx 10^{-66+2n}$, where
$10^{-n}$ is the measuring scale and the Planck scale $\sim
10^{-33}cm$ is assumed. The second one is also too small for the
conventional temperatures, that is those much below the Planck's.
It is demonstrated that relative to the first of these parameters
the Universe may be considered as a nonuniform lattice in the
four-dimensional hypercube with dimensionless finite-length (1/4)
edges. And the time variable is also described by one of the
above-mentioned dimensions due to the second parameter and
generalized uncertainty relations in thermodynamics. In this
context the lattice is understood as a deformation rather than
approximation.
\end{abstract}

\section{Introduction}
In the last decades the scientists have become aware that Quantum
Mechanics of the early Universe should be different from the
classical Quantum Mechanics. To illustrate, in the first the
Generalized Uncertainty Relations (GUR)\cite{r1},\cite{r2} are
valid, whereas in the second one the ordinary Uncertainty
Relations (UR) of Heisenberg \cite{r3}are effective. Resultant
from GUR is the fundamental length on the order of Planck's
\cite{r4} that is lacking in Quantum Mechanics with UR. Thus,
GUR-involving Quantum Mechanics may be considered as a deformation
of Quantum Mechanics with UR or in other words Quantum Mechanics
with Fundamental Length(QMFL) is a deformation of well-known
Quantum Mechanics. The "deformation" is understood as a theory
extension by inclusion of one or several parameters so that the
original theory be associated with the limit, where the indicated
parameters are tending to some fixed values \cite{r5}. QM being a
deformation of the Classical Mechanics presents a vivid example.
The deformation parameter in this case is Planck's constant
$\hbar$. When $\hbar\rightarrow 0$, QM is transformed to the
Classical Mechanics. The deformation in Quantum Mechanics at
Planck scale takes different paths: commutator deformation or more
precisely deformation of the respective Heisenberg algebra
\cite{r6},\cite{r7},\cite{r8} and the density matrix deformation
approach \cite{r9}--\cite{r15}. The first approach suffers from
two serious disadvantages: 1) the deformation parameter is a
dimensional variable $\kappa$ with a dimension of mass \cite{r6};
2) in the limiting transition to QM this parameter goes to
infinity and fluctuations of other values are hardly sensitive to
it. The second approach is devoid of such limitations as in this
approach the deformation parameter is represented by the
dimensionless quantity $\alpha=l_{min}^{2}/x^{2}$, where $x$ is
the measuring scale and the variation interval $\alpha$ is finite
$0<\alpha\leq1/4$ \cite{r9}--\cite{r13}. Moreover, it gives a key
to the solution of particular problems: an extra term in Liouville
equation in the processes associated with black holes
\cite{r11}--\cite{r13}; singularities and cosmic censorship
\cite{r12},\cite{r13};derivation of a semiclassical
Bekenstein-Hawking formula for the black hole entropy and some
others \cite{r13},\cite{r16}. In \cite{r16}--\cite{r18}it has been
shown that within the developing paradigm there is a possibility
for the solution of Hawking's information Paradox problem
\cite{r19}--\cite{r21}and $\alpha$ may be interpreted as a new
small parameter of quantum theory. Besides, over the dimensionless
interval $I_{1/4}=(0;1/4]$this parameter takes on a series of
discrete values nonuniformly filling the indicated interval as
distinct from the conventional lattice. This lattice may be
considered in the ordinary cube, each edge of which is associated
with a particular space dimension. This property of $\alpha$ will
be considered further together with similar feature for its
counterpart in Statistical Mechanics - $\tau,\tau\in I_{1/4}$
parameter. It will be demonstrated that due to the latter and
Generalized Uncertainty Relations in thermodynamics the time
variable may be also treated as a discrete and nonuniform series
over the same interval $I_{1/4}=(0;1/4]$. Thus, any theory may be
considered as a nonuniform lattice in the four-dimensional
hypercube $I_{1/4}^{4}$. In this context the lattice is understood
as a deformation rather than approximation.

\section {Relevant Suggestions and Refinements}

In this section we recall in short the introduction of $\alpha$
and $\tau$ parameters into the generalized Quantum and Statistical
Mechanics. Here "generalized" means the Quantum and Statistical
Mechanics describing the processes both in the current
(conventional scales)and early (scales on the order of Planck's)
Universe. Now it is obvious that in the latter the notion of the
fundamental (minimum) length $l_{min}\sim l_{p}$ is a requisite
\cite{r4}, where $l_{p}$ is the Planck's length. It has been
demonstrated \cite{r9}--\cite{r13} that on retention of a
well-known measuring procedure the density matrix becomes
dependent on the additional parameter $\alpha=l_{min}^{2}/x^{2}$,
where $x$ is the measuring scale. In this way the density matrix
is subjected to the deformation procedure (e.g.\cite{r5}), the
resultant deformed object is referred to as {\bf density
pro-matrix}, whereas the conventional density matrix and exact
definition will be as follows: \cite{r9}--\cite{r13}:
\\
\noindent {\bf Definition 1.} {\bf(Quantum Mechanics with
Fundamental Length [for Neumann's picture])}
\\
\\
\noindent Any system in QMFL is described by a density pro-matrix
of the form $${\bf \rho(\alpha)=\sum_{i}\omega_{i}(\alpha)|i><i|},$$
where
\begin{enumerate}
\item Vectors $|i>$ form a full orthonormal system;
\item $\omega_{i}(\alpha)\geq 0$ and for all $i$  the
finite limit $\lim\limits_{\alpha\rightarrow
0}\omega_{i}(\alpha)=\omega_{i}$ exists;
\item
$Sp[\rho(\alpha)]=\sum_{i}\omega_{i}(\alpha)<1$,
$\sum_{i}\omega_{i}=1.$;
\item For every operator $B$ and any $\alpha$ there is a
mean operator $B$ depending on  $\alpha$:\\
$$<B>_{\alpha}=\sum_{i}\omega_{i}(\alpha)<i|B|i>.$$
\item The following condition should be fulfilled:
\begin{equation}\label{U1}
Sp[\rho(\alpha)]-Sp^{2}[\rho(\alpha)]\approx\alpha.
\end{equation}
Consequently we can find the value for $Sp[\rho(\alpha)]$
satisfying the above-stated condition:
\begin{equation}\label{U2}
Sp[\rho(\alpha)]\approx\frac{1}{2}+\sqrt{\frac{1}{4}-\alpha}
\end{equation}
and therefore
\item $0<\alpha\leq1/4$.
\end{enumerate}

It is no use to enumerate all the evident implications and
applications of {\bf Definition 1.}, better refer to
\cite{r12},\cite{r13}. Nevertheless, it is clear that
\\ {\bf for $\alpha\rightarrow 0$ the above limit covers
both the Classical or Quantum Mechanics depending on
$\hbar\rightarrow 0$ or not}.
\\It should be noted that according to
{\bf Definition 1.} a minimum measurable length is equal
 to $l^{*}_{min}=2l_{min}$ being a nonreal
number at point $l_{min}$,$Sp[\rho(\alpha)]$. Because of this, a
space part of the Universe is a lattice with a spacing of
$a_{min}=2l_{min}\sim 2l_{p}$. In consequence the first issue concerns
the lattice spacing of any lattice-type model(for example
\cite{rLat1,rLat2}): a selected lattice spacing $a_{lat}$
should not be less than $a_{min}$,i.e. always $a_{lat}\geq a_{min}>0$.
Besides, a continuum limit in any lattice-type model is meaning
$a_{lat}\rightarrow a_{min}>0$ rather than $a_{lat}\rightarrow 0$.
\\ Proceeding from $\alpha$, for each space dimension we have a
discrete series of rational values for the inverse squares of even
numbers nonuniformly distributed along the real number line
$\alpha\div 1/4, 1/16,1/36,1/64,...$. A question arises,is this
series somewhere terminated or, on the contrary, is it infinite?
The answer depends on the answers to two other questions:
(1)is there theoretically a maximum measurability limit for the
scales $l_{max}$?; and (2) is our Universe closed in the sense that
its extension may be sometime replaced by compression, when a maximum
extension precisely gives a maximum scale $l_{max}$?  Should an
answer to one of these questions be positive, we should have
condition 6 {\bf Definition 1.} rather than
$0<l^{2}_{min}/l^{2}_{max}\leq\alpha\leq1/4$.
\\ Note that in the majority of cases all three space dimensions are
equal, at least at large scales, and hence their associated values of
$\alpha$ parameter should be identical. This means that for most
cases, at any rate in the large-scale (low-energy) limit, a single
deformation parameter $\alpha$ is sufficient to accept one and the same
value for all three dimensions to a high degree of accuracy. In the
general case, however, this is not true, at least for very high energies
(on the order of the Planck's), i.e. at Planck scales, due to
noncommutativity of the spatial coordinates \cite{r1},\cite{r2},\cite{r6}:
\\
$$\left[ x_i ,x_j \right]\neq 0$$
\\
In consequence in the general case we have a point with coordinates
${\bf \widetilde{\alpha}}=(\alpha_{1},\alpha_{2},\alpha_{3})$
in the normal(three-dimensional) cube $I_{1/4}^{3}$ of side
$I_{1/4}=(0;1/4]$.
\\ It should be noted that this universal cube may be extended to
the four-dimensional hypercube by inclusion of the additional parameter
$\tau,\tau\in I_{1/4}$ that is generated by internal energy of the
statistical ensemble and its temperature for the events when this notion
is the case. It will be recalled that $\tau$ parameter occurs from a
maximum temperature that is  in its turn generated by the Generalized
Uncertainty Relations of "energy – time" pair in GUR.
The exact definition \cite{r14},\cite{r15}is as follows:
\\
\noindent {\bf Definition 2.} {\bf(Deformation of Statistical
Mechanics)} \noindent \\Deformation of Gibbs distribution valid
for temperatures on the order of the Planck's $T_{p}$ is described
by deformation of a statistical density matrix
(statistical density pro-matrix) of the form
\\$${\bf \rho_{stat}(\tau)=\sum_{n}\omega_{n}(\tau)|\varphi_{n}><\varphi_{n}|}$$
having the deformation parameter
$\tau = T^{2}/T_{max}^{2}$, where
\begin{enumerate}
\item The vectors $|\varphi_{n}>$ form a full orthonormal system;
\item $\omega_{n}(\tau)\geq 0$ and for all $n$ at $\tau \ll 1$
 we obtain
 $\omega_{n}(\tau)\approx \omega_{n}=\frac{1}{Q}\exp(-\beta E_{n})$
In particular, $\lim\limits_{T_{max}\rightarrow \infty
(\tau\rightarrow 0)}\omega_{n}(\tau)=\omega_{n}$
\item
$Sp[\rho_{stat}(\tau)]=\sum_{n}\omega_{n}(\tau)<1$,
$\sum_{n}\omega_{n}=1$;
\item For every operator $B$ and any $\tau$ there is a
mean operator $B$ depending on  $\tau$ \\
$$<B>_{\tau}=\sum_{n}\omega_{n}(\tau)<n|B|n>.$$
\item  Finally,  the following condition must be fulfilled:
\begin{equation}\label{U12b}
Sp[\rho_{stat}(\tau)]-Sp^{2}[\rho_{stat}(\tau)]\approx \tau.
\end{equation}
Hence we can find the value for $Sp[\rho_{stat}(\tau)]$
satisfying the condition of Definition 2 (similar to Definition 1):
\begin{equation}\label{U13}
Sp[\rho_{stat}(\tau)]\approx\frac{1}{2}+\sqrt{\frac{1}{4}-\tau}.
\end{equation}
This implies that
\item $0<\tau \leq 1/4$
\end{enumerate}
So $\tau$ is a counterpart (twin) of $\alpha$, yet for the Statistical
Mechanics. At the same time, originally for $\tau$ nothing implies
the discrete properties of parameter $\alpha$ indicated above:
\\ for $\tau$ there is a discrete series (lattice) of the rational
values of inverse squares for even numbers not uniformly distributed
along the real number line: $\tau\div 1/4, 1/16, 1/36,1/64,...$.
 \\ Provided such a series exists actually,
\\** the finitness and infinity question for this series amounts to
two other questions:
\\ (1) is there theoretically any minimum measurability limit for
the average temperature of the Universe $T_{min}\neq 0$ and
(2)is our Universe closed in a sense that its extension may be
sometime replaced by compression? Then maximum extension just gives a
minimum temperature $T_{min}\neq 0$.
\\ The question concerning the discretization of parameter $\tau$
is far from being idle. The point is that originally by its nature
this parameter seems to be continuous as it is associated with
temperature. Nevertheless, in the following section we show that actually
$\tau$ is dual in nature: it is directly related to time that is in turn
quantized in the end giving a series $\tau\div 1/4, 1/16, 1/36,1/64,...$.

\section{Dual Nature of Parameter $\tau$ and its Temporal Aspect}

In this way when at point ${\bf\widetilde{\alpha}}$ of the normal
(three-dimensional) cube $I_{1/4}^{3}$ of side $I_{1/4}=(0;1/4]$ an
additional "temperature" variable $\tau$ is added, a nonuniform lattice
of the point results, where we denote
$\widetilde{\alpha}_{\tau}=(\widetilde{\alpha},\tau)=
(\alpha_{1},\alpha_{2},\alpha_{3},\tau)$ at the four-dimensional
hypercube $I_{1/4}^{4}$, every coordinate of which assumes one and
the same discrete series of values: 1/4, 1/16, 1/36,1/64,...,
$1/4n^{2}$,... .(Further it is demonstrated that $\tau$ is also taking
on a discrete series of values.) The question arises, whether time
"falls" within this picture. The answer is positive. Indeed, parameter
$\tau$ is dual (thermal and temporal) in nature owing to introduction
of the Generalized Uncertainty Relations in Thermodynamics (GURT)
\cite{r15},\cite{r22},\cite{r23}:
\\
$$\Delta \frac{1}{T} \geq
  \frac{k}{\Delta U}+\alpha^{\prime}
  \frac{1}{T_{p}^2}\,
  \frac{\Delta U}{k}+...,$$
\\
where $k$ - Boltzmann constant, $T$ - temperature of the ensemble, $U$ -
its internal energy. A direct implication of the latter inequality
is occurrence of a "maximum" temperature $T_{max}$ that is inversely
proportional to "minimal" time
$t_{min}\sim t_{p}$ (\cite{r15} expression (11)):
\\
$$T_{max}=\frac{\hbar}{2\surd \alpha^{\prime}t_{p}
k}=\frac{\hbar}{\Delta t_{min} k}$$
\\
However, $t_{min}$ follows from the Generalized Uncertainty Relations
in Quantum Mechanics for "energy-time" pair
\cite{r14},\cite{r15},\cite{r22},\cite{r23}:
\\
$$\Delta t\geq\frac{\hbar}{\Delta
E}+\alpha^{\prime}t_{p}^2\,\frac{\Delta E}{ \hbar}$$
\\
Thus, $T_{max}$ is the value relating GUR and GURT together \cite{r15},\cite{r22},\cite{r23}:
\begin{equation}\label{U18}
\left\{
\begin{array}{lll}
\Delta x & \geq & \frac{\displaystyle\hbar}{\displaystyle\Delta
p}+\alpha^{\prime} L_{p}^2\,\frac{\displaystyle\Delta
p}{\displaystyle\hbar}+... \\
  &  &  \\
  \Delta t & \geq &  \frac{\displaystyle\hbar}{\displaystyle\Delta E}+\alpha^{\prime}
  t_{p}^2\,\frac{\displaystyle\Delta E}{\displaystyle\hbar}+... \\
  &  &  \\

  \Delta \frac{\displaystyle 1}{\displaystyle T} & \geq &
  \frac{\displaystyle k}{\displaystyle\Delta U}+\alpha^{\prime}
  \frac{\displaystyle 1}{\displaystyle T_{p}^2}\,
  \frac{\displaystyle\Delta U}{\displaystyle k}+...,
\end{array} \right.
\end{equation},
since the thermodinamical value $T_{max}$ (GURT) is associated with
the quantum-mechanical one $E_{max}$ (GUR) by the formula
\cite{r14},\cite{r15},\cite{r22},\cite{r23}:
\\
$$T_{max}=\frac{E_{max}}{k}$$
\\
The notion of value $t_{min}\sim 1/T_{max}$ is physically crystal clear,
it means a minimum time for which any variations in the energy spectrum
of every physical system may be recorded. Actually, this value is equal
to $t^{*}_{min}=2t_{min}\sim t_{p}$ as at the initial points $l_{min}$
and $T_{max}$ the spurs of the quantum-mechanical and statistical density
pro-matrices ${\bf\rho_(\alpha)}$ and ${\bf\rho_{stat}(\tau)}$
are complex, determined only beginning from
$2l_{min}$ è $T^{*}_{max}=\frac{1}{2}T_{max}$
\cite{r13}--\cite{r15}that is associated with the same time point
$t^{*}_{min}=2t_{min}$. For QMFL this has been noted in the previous
section.
\\  In such a manner a discrete series $l^{*}_{min},2l^{*}_{min}$,...
generates in QMFL the discrete time series
$t^{*}_{min},2t^{*}_{min},...$, that is in turn associated (due to
GURT)with a discrete temperature series
$T^{*}_{max}$,$\frac{1}{2}T^{*}_{max}$, ... . From this it is
inferred that a "temperature" scale $\tau$ may be interpreted as a
"temporal" one $\tau=t_{min}^{2}/t^{2}$. In both cases the
generated series has one and the same discrete set of values of
parameter $\tau$ :$\tau\div 1/4, 1/16, 1/36,1/64,...,
1/4n^{2}$,... . Thus, owing to time quantization in QMFL one is
enabled to realize quantization of temperature in the generalized
Statistical Mechanics with the use of GURT.
\\  Using $Lat_{\widetilde{\alpha}}$ we denote the lattice in cube $I_{1/4}^{3}$ formed by points $\widetilde{\alpha}$, and through
$Lat^{\tau}_{\widetilde{\alpha}}$ we denote the lattice in hypercube
$I_{1/4}^{4}$, that is formed by points
$\widetilde{\alpha}_{\tau}=(\widetilde{\alpha},\tau)$.

\section{Quantum Theory \\ for the Lattice in Hypercube}

Any quantum theory may be defined for the indicated lattice in hypercube.
To this end it is required to go from Neumann's picture to
Shr{\"o}dinger's picture. We recall the fundamental definition
\cite{r13},\cite{r17},\cite{r18} with $\alpha$ changed by $\widetilde{\alpha}$:
\\ \noindent {\bf Definition
$1^{\prime}$} {\bf Quantum Mechanics with Fundamental Length}
\\ {\bf (Shr{\"o}dinger's picture)}
\\
Here, the prototype of Quantum Mechanical normed wave function (or
the pure state prototype) $\psi(q)$ with $\int|\psi(q)|^{2}dq=1$
in QMFL is
$\psi(\widetilde{\alpha},q)=\theta(\widetilde{\alpha})\psi(q)$.
The parameter of deformation $\widetilde{\alpha}\in I_{1/4}^{3}$.
Its properties are
$|\theta(\widetilde{\alpha})|^{2}<1$,$\lim\limits_{|\widetilde{\alpha}|\rightarrow
0}|\theta(\widetilde{\alpha})|^{2}=1$ and the relation
$|\theta(\alpha_{i})|^{2}-|\theta(\alpha_{i})|^{4}\approx
\alpha_{i}$ takes place. In such a way the total probability
always is less than 1:
$p(\widetilde{\alpha})=|\theta(\widetilde{\alpha})|^{2}
=\int|\theta(\widetilde{\alpha})|^{2}|\psi(q)|^{2}dq<1$ tending to
1, when  $\|\widetilde{\alpha}\|\rightarrow 0$. In the most
general case of the arbitrarily normed state in QMFL(mixed state
prototype)
$\psi=\psi(\widetilde{\alpha},q)=\sum_{n}a_{n}\theta_{n}(\widetilde{\alpha})\psi_{n}(q)$
with $\sum_{n}|a_{n}|^{2}=1$ the total probability is
$p(\widetilde{\alpha})=\sum_{n}|a_{n}|^{2}|\theta_{n}(\widetilde{\alpha})|^{2}<1$
and
 $\lim\limits_{\|\widetilde{\alpha}\|\rightarrow 0}p(\widetilde{\alpha})=1$.

It is natural that Shr{\"o}dinger equation is also deformed in
QMFL. It is replaced by the equation

\begin{equation}\label{U24S}
\frac{\partial\psi(\widetilde{\alpha},q)}{\partial t}
=\frac{\partial[\theta(\widetilde{\alpha})\psi(q)]}{\partial
t}=\frac{\partial\theta(\widetilde{\alpha})}{\partial
t}\psi(q)+\theta(\widetilde{\alpha})\frac{\partial\psi(q)}{\partial
t},
\end{equation}
where the second term in the right-hand side generates the
Shr{\"o}dinger equation as
\begin{equation}\label{U25S}
\theta(\widetilde{\alpha})\frac{\partial\psi(q)}{\partial
t}=\frac{-i\theta(\widetilde{\alpha})}{\hbar}H\psi(q).
\end{equation}
Here $H$ is the Hamiltonian and the first member is added
similarly to the member that appears in the deformed Liouville
equation, vanishing when $\theta[\widetilde{\alpha}(t)]\approx
const$. In particular, this takes place in the low energy limit in
QM, when $\|\widetilde{\alpha}\|\rightarrow 0$. It should be noted
that the above theory is not a time reversal of QM because the
combination $\theta(\widetilde{\alpha})\psi(q)$ breaks down this
property in the deformed Shr{\"o}dinger equation. Time-reversal is
conserved only in the low energy limit, when a quantum mechanical
Shr{\"o}dinger equation is valid.
\\ According to {\bf Definition $1^{\prime}$}everywhere $q$ is
the coordinate of point at the three-dimensional space. As indicated
in \cite{r9}--\cite{r18},
for a density pro-matrix there exists an exponential ansatz
satisfying the formula  \ref{U1} â {\bf Definition 1}:
\begin{equation}\label{U26S}
\rho^{*}(\alpha)=\sum_{i}\omega_{i} exp(-\alpha)|i><i|,
\end{equation}
where all $\omega_{i}>0$ are independent of $\alpha$  and their
sum is equal to 1. In this way
$Sp[\rho^{*}(\alpha)]=exp(-\alpha)$. Then in the momentum
representation $\alpha=p^{2}/p^{2}_{max}$, $p_{max}\sim
p_{pl}$,where $p_{pl}$ is the Planck momentum. When present in
matrix elements, $exp(-\alpha)$  damps the contribution of great
momenta in a perturbation theory.
\\ It is clear that for each of the coordinates $q_{i}$ the
exponential ansatz makes a contribution to the deformed wave function
$\psi(\widetilde{\alpha},q)$ the modulus of which equals
$exp(-\alpha_{i}/2)$  and, obviously, the same contribution
to the conjugate function
$\psi^{*}(\widetilde{\alpha},q)$. Because of this, for
exponential ansatz one may write
\begin{equation}\label{U27S}
\psi(\widetilde{\alpha},q)=\theta(\widetilde{\alpha})\psi(q),
\end{equation}
where $|\theta(\widetilde{\alpha})|=exp(-\sum_{i}\alpha_{i}/2)$.
As noted above, the last exponent of the momentum representation
reads $exp(-\sum_{i}p_{i}^{2}/2p_{max}^{2})$ and in this way
it removes UV (ultra-violet) divergences in the theory.

It follows that $\widetilde{\alpha}$ is a new small parameter.
Among its obvious advantages one could name:
\\1)  its dimensionless nature,
\\2)  its variability over the finite interval $0<\alpha_{i}\leq 1/4$.
Besides, for the well-known physics it is actually very small:
$\alpha\sim 10^{-66+2n}$, where $10^{-n}$ is the measuring scale.
Here the Planck scale $\sim 10^{-33}cm$ is assumed;
\\3)and finally the calculation of this parameter involves all
three fundamental constants, since by Definition 1 of section 2
$\alpha_{i}= l_{min}^{2}/x_{i}^{2 }$, where $x_{i}$ is the
measuring scale on i-coordinate and $l_{min}^{2}\sim
l_{pl}^{2}=G\hbar/c^{3}$.
\\ Therefore, series expansion in $\alpha_{i}$ may be of great importance.
Since all the field components and hence the Lagrangian will be
dependent on $\widetilde{\alpha}$, i.e. $\psi=\psi(\widetilde{\alpha}),L=L(\widetilde{\alpha})$, quantum theory
may be considered as a theory of lattice $Lat_{\widetilde{\alpha}}$ and
hence of lattice $Lat^{\tau}_{\widetilde{\alpha}}$.

\section{Introduction of Quantum Field Theory and Initial Analysis}
With the use of this approach for the customary energies a Quantum
Field Theory (QFT) is introduced with a high degree of accuracy. In our
context "customary" means the energies much lower than the Planck ones.
\\ It is important that as the spacing of lattice $Lat^{\tau}_{\widetilde{\alpha}}$ is decreasing in inverse proportion to
the square of the respective node, for a fairly large node number
$N>N_{0}$  the lattice edge beginning at this node
$\ell_{N,N+1}$ \cite{r9}--\cite{r13}will be of length
$\ell_{N,N+1}\sim 1/4N^{3}$, and by this means edge lengths of the lattice
are rapidly decreasing with the spacing number. Note that in the large-scale
limit this (within any preset accuracy)leads to parameter
$\alpha=0$, pure states and in the end to QFT. In this way a theory for
the above-described lattice presents a deformation of the originally
continuous variant of this theory as within the developed approach continuity
is accurate to $\approx 10^{-66+2n}$, where $10^{-n}$ is the measuring scale
and
the Planck scale $\sim 10^{-33}cm$ is assumed. Whereas the lattice per se $Lat^{\tau}_{\widetilde{\alpha}}$ may be interpreted as a deformation of
the space continuum with the deformation parameter equal to the
varying edge length $\ell_{\alpha^{1}_{\tau_{1}},\alpha^{2}_{\tau_{2}}}$,
where $\alpha^{1}_{\tau_{1}}$ è $\alpha^{2}_{\tau_{2}}$ are two adjacent
points of the lattice $Lat^{\tau}_{\widetilde{\alpha}}$. Proceeding from this, all well-known theories including $\varphi^{4}$, QED, QCD and so on may be
studied based on the above-described lattice.
\\ Here it is expedient to make the following remarks:
\\{\bf (1) going on from the well-known energies of these theories to higher energies (UV behavior) means a change from description of the theory's
behavior for the lattice portion with high edge numbers to the portion
with low numbers of the edges;
\\ (2) finding of quantum correction factors for the primary deformation
parameter $\widetilde{\alpha}$ is a power series expansion in each
$\alpha_{i}$. In particular, in the simplest case (Definition
$1^{\prime}$)means expansion of the left side in relation
$|\theta(\alpha_{i})|^{2}-|\theta(\alpha_{i})|^{4}\approx
\alpha_{i}$:
\\
$$|\theta(\alpha_{i})|^{2}-|\theta(\alpha_{i})|^{4}=\alpha_{i}+a_{0}\alpha^{2}_{i}+a_{1}\alpha^{3}_{i}+...$$
\\
and calculation of the associated coefficients $a_{0},a_{1},...$.}

This approach to calculation of the quantum correction factors may be
used in the formalism for density pro-matrix (Definition 1).  In this case,
the primary relation \ref{U1} may be written in the form of a series
\begin{equation}\label{U1b}
Sp[\rho(\alpha)]-Sp^{2}[\rho(\alpha)]=\alpha+a_{0}\alpha^{2}
+a_{1}\alpha^{3}+...
\end{equation}
As a result, a measurement procedure using the exponential ansatz
(\ref{U26S}) may be understood as the calculation of factors
$a_{0}$,$a_{1}$,... or the definition of additional members in the
exponent "destroying" $a_{0}$,$a_{1}$,... \cite{r14},\cite{r18}.
It is easy to check that the exponential ansatz gives
$a_{0}=-3/2$, being coincident with the logarithmic correction
factor for the Black Hole entropy \cite{r24}.
\\ Most often a quantum theory is considered at zero temperature
$T=0$, in this context amounting to nesting of the
three-dimensional lattice $Lat_{\widetilde{\alpha}}$ into the
four-dimensional one:
$Lat^{\tau}_{\widetilde{\alpha}}$:$Lat_{\widetilde{\alpha}}\subset
Lat^{\tau}_{\widetilde{\alpha}}$ and nesting of the cube
$I_{1/4}^{3}$ into the hypercube $I_{1/4}^{4}$ as a bound given by
equation $\tau=0$. However, in the most general case the points
with nonzero values of $\tau$ may be important as there is a
possibility for nonzero temperature $T\neq0$ (quantum field theory
at finite temperature) that is related to the value of $\tau$
parameter, though very small but still nonzero: $\tau\neq0$. To
illustrate: in QCD for the normal lattice \cite{r25} a critical
temperature $T_{c}$ exists so that the following is fulfilled:
\\
at $$T<T_{c}$$ the confinement phase occurs,
\\
and for $$T>T_{c}$$ the deconfinement is the case.
\\ A critical temperature $T_{c}$ is associated with the "critical"
parameter $\tau_{c}=T^{2}_{c}/T^{2}_{max}$  and the selected bound
of hypercube $I_{1/4}^{4}$ set by equation $\tau=\tau_{c}>0$.

\section{Conclusion}
The principal issue of the present work is the development of a
unified approach to study all the available quantum theories
without exception owing to the proposed small dimensionless
parameter – deformation parameter: $\widetilde{\alpha}_{\tau}\in
Lat^{\tau}_{\widetilde{\alpha}}$ that is in turn dependent on all
the fundamental constants $G,c,\hbar$ and  $k$.
\\ Thus, there is a reason to believe that lattices $Lat_{\widetilde{\alpha}}$
and $Lat^{\tau}_{\widetilde{\alpha}}$ may be a universal means to
study different quantum theories. This poses a number of intriguing
problems:
\\ (1)description of a set of lattice symmetries $Lat_{\widetilde{\alpha}}$
and $Lat^{\tau}_{\widetilde{\alpha}}$.
\\ (2) for each of the well-known physical theories
($\varphi^{4}$,QED,QCD and so on) definition of the selected (special)
points (phase transitions, different symmetry violations, etc.)
associated with the above-mentioned lattices.
\\ These problems of current importance necessitate further investigation by the author.

%References


\begin{thebibliography}{99}
%
%
\bibitem{r1}
D. V. Ahluwalia,Quantum Measurement, Gravitation, and Locality,
Phys.Lett. B339 (1994)301;A.Kempf, G.Mangano, R.B.Mann. Hilbert
Space Representation of the Minimal Length Uncertainty Relation.
Phys.Rev. D52 (1995)1108.
%
%
\bibitem{r2}
D.V.Ahluwalia,Wave-Particle duality at the Planck scale: Freezing
of neutrino oscillations, Phys.Lett. A275 (2000)31;Interface of
Gravitational and Quantum Realms, Mod.Phys.Lett. A17(2002)1135
%
%
\bibitem{r3}
W.Heisenberg. Uber den anschaulichen Inhalt der
quantentheoretischen Kinematik und Mechanik.
 Zeitsch.fur Phys. {\bf 43}, 172-184, (1927).
%
%
\bibitem{r4}
R.J.Adler and D.I.Santiago. On Gravity and the Uncertainty
Principle. Mod.Phys.Lett. {\bf A14}, 1371-1377 (1999).
%
%
\bibitem{r5}
L.Faddeev. Mathematical View on Evolution of Physics. Priroda.
{\bf 5}, 11-18, (1989).
%
%
\bibitem{r6}
M.Maggiore, The algebraic structure of the generalized uncertainty
principle,Phys.Lett.B319(1993)83
%
%
\bibitem{r7}
M.Maggiore,Quantum Groups,Gravity and Generalized Uncertainty
Principle, Phys.Rev.D49(1994)5182
%
%
\bibitem{r8}
S.Capozziello,G.Lambiase and G.Scarpetta, The Generalized
Uncertainty Principle from Quantum Geometry, Int.J.Theor.Phys. 39
(2000),15
%
%
\bibitem{r9}
A.E.Shalyt-Margolin, Fundamental Length,Deformed Density Matrix
and New View on the Black Hole Information Paradox,[gr-qc/0207074]
%
%
\bibitem{r10}
A.E.Shalyt-Margolin and A.Ya.Tregubovich, Generalized Uncertainty
Relations,Fundamental Length and Density Matrix,[gr-qc/0207068]
%
%
\bibitem{r11}
A.E.Shalyt-Margolin and J.G.Suarez, Density Matrix and Dynamical
aspects of Quantum Mechanics with Fundamental Length,
[gr-qc/0211083]
%
%
\bibitem{r12}
A.E.Shalyt-Margolin and J.G.Suarez,Quantum Mechanics of the Early
Universe and its Limiting Transition,[gr-qc/0302119]
%
%
\bibitem{r13}
A.E.Shalyt-Margolin and J.G.Suarez,Quantum Mechanics at Planck's
scale and Density Matrix,Intern.Journ.of Mod.Phys.D.12(2003)1265
%
%
\bibitem{r14}
A.E.Shalyt-Margolin,Density Matrix in Quantum and Statistical
Mechanics at Planck-Scale [gr-qc/0307057]
%
%
\bibitem{r15}
A.E.Shalyt-Margolin,A.Ya.Tregubovich,Deformed Density Matrix and
Generalized Uncertainty Relation in  Thermodynamics,Mod. Phys.
Lett. A, Vol.19, No.1(2004)pp.71-81,[hep-th/0311034].
%
%
\bibitem{r16}
A.E.Shalyt-Margolin. Deformed density matrix, Density of entropy
and Information problem,[gr-qc/0307096].
%
%
\bibitem{r17}
A.E.Shalyt-Margolin, Non-Unitary and Unitary Transitions in
Generalized Quantum  Mechanics and Information Problem Solving,
[hep-th/0309121]
%
%
\bibitem{r18}
A.E.Shalyt-Margolin,Non-Unitary and Unitary Transitions in
Generalized Quantum  Mechanics, New Small Parameter and
Information Problem Solving,(to be published in Mod.Phys.Lett.
A),[hep-th/0311239].
%

%
\bibitem{r19}
S.Hawking,Breakdown of Predictability in Gravitational Collapse,
Phys.Rev.D14(1976)2460
%
%
\bibitem{r20}
S.Giddings,The Black Hole Information Paradox,[hep-th/9508151]
%
%
\bibitem{r21}
A.Strominger, Les Houches Lectures on Black Holes,
[hep-th/9501071].
%
%
\bibitem{rLat1}
H.Grosse,Models in Statistical Physics and Quantum Field Theory,
Springer-Verlag,1988
%
%
\bibitem{rLat2}
C.Itzykson,J-M.Drouffe,Statistical Field Theory,Vol.1,2.,
\\Cambridge University Press,Cambridge,1991.
%
%
\bibitem{r22}
A.E.Shalyt-Margolin and A.Ya.Tregubovich, Generalized Uncertainty
Relations in a Quantum Theory and Thermodynamics From the Uniform
Point of View [gr-qc/0204078]
%
%
\bibitem{r23}
A.E.Shalyt-Margolin and A.Ya.Tregubovich, Generalized Uncertainty
Relations in  Thermodynamics [gr-qc/0307018]
%
%
\bibitem{r24}
P.Majumdar,  Black hole entropy: classical and quantum aspects,
{Expanded version of lectures given at the YATI Conference on
Black Hole Astrophysics, Kolkata, India, April 2001},
[hep-th/0110198]; S.Das, P.Majumdar and  R.K. Bhaduri, General
Logarithmic Corrections to Black Hole Entropy,hep-th/0111001;
E.C.Vagenas, Semiclassical  Corrections to the Bekenstein-Hawking
Entropy of the BTZ Black Hole via
Selfgravitation,Phys.Lett.B533(2002)302, [hep-th/0109108].
%
%
\bibitem{r25}
Adriano Di Giacommo, Confinement of Color: A Review,Talk at XII
International Conference on Selected Problems of Modern Physics,
Dubna, June 2003,[hep-lat/0310023].
%
%
\end{thebibliography}
\end{document}